\PassOptionsToPackage{table}{xcolor}

\documentclass[11pt,a4paper]{article}
\usepackage[hyperref]{ranlp2025}
\usepackage{times}
\usepackage{latexsym}

\usepackage{microtype}

\usepackage{xcolor}
\usepackage{graphicx}
\usepackage{multirow}
\usepackage{url}
\usepackage{booktabs}
\usepackage[inline]{enumitem}
\usepackage{caption}
\usepackage{subcaption}
\usepackage{algorithm}
\usepackage{algpseudocode}
\usepackage{amsmath}
\usepackage{makecell}

\aclfinalcopy 

\title{LLM-based Embedders for Prior Case Retrieval}

\author{Damith Premasiri, Tharindu Ranasinghe, \textbf{Ruslan Mitkov} \\
 School of Computing and Communications, Lancaster University, UK \\ {\tt  \{d.dolamullage, t.ranasinghe, r.mitkov\}@lancaster.ac.uk} }

\date{}

\begin{document}
\maketitle
\begin{abstract}
In common law systems, legal professionals such as lawyers and judges rely on precedents to build their arguments. As the volume of cases has grown massively over time, effectively retrieving prior cases has become essential. Prior case retrieval (PCR) is an information retrieval (IR) task that aims to automatically identify the most relevant court cases for a specific query from a large pool of potential candidates. While IR methods have seen several paradigm shifts over the last few years, the vast majority of PCR methods continue to rely on traditional IR methods, such as \texttt{BM25}. The state-of-the-art deep learning IR methods have not been successful in PCR due to two key challenges: \begin{enumerate*}[label={\textbf{\textit{\roman*}}.}] \item Lengthy legal text limitation; when using the powerful BERT-based transformer models, there is a limit of input text lengths, which inevitably requires to shorten the input via truncation or division with a loss of legal context information. \item Lack of legal training data; due to data privacy concerns, available PCR datasets are often limited in size, making it difficult to train deep learning-based models effectively \end{enumerate*}. In this research, we address these challenges by leveraging LLM-based text embedders in PCR. LLM-based embedders support longer input lengths, and since we use them in an unsupervised manner, they do not require training data, addressing both challenges simultaneously. In this paper, we evaluate state-of-the-art LLM-based text embedders in \textbf{four} PCR benchmark datasets and show that they outperform \texttt{BM25} and supervised transformer-based models. 
\end{abstract}

\section{Introduction}
Information retrieval (IR) systems have progressed through several paradigm shifts in the last few decades \cite{zhu2023large, plum-etal-2024-guided}. Initial IR methods relied on term-based methods such as \texttt{BM25} \cite{robertson2009probabilistic} and boolean logic, focusing on keyword matching for document retrieval \cite{chowdhury2010introduction}. The paradigm gradually shifted with the introduction of vector space models, enabling a more sophisticated understanding of the semantic relationships between queries and documents \cite{10.1145/361219.361220}. Initially, these models relied on statistical language models \cite{10.1145/319950.320022}, but in recent years, neural vector space models have achieved remarkable performance in IR \cite{10.1145,xiong2021approximate}. More recently, large language models (LLMs) have been integrated into these vector space models as embedders, further improving the performance \cite{10.1145/3626772.3657951, musts}. LLM-based embedders have dominated text retrieval benchmarks such as \texttt{MTEB} \cite{muennighoff-etal-2023-mteb} and \texttt{BEIR} \cite{thakur2021beir}. 

Prior case retrieval (PCR) is an IR application where the goal is to retrieve cases from a large legal database of historical cases that are similar to a given query case \cite{10.1007/978-3-031-15931-2_47,feng-etal-2024-legal,10.1145/3459637.3481994,Tran2020}. PCR holds substantial practical value, regardless of the legal system a country follows. In the Common Law System (as in the United Kingdom and India), legal professionals, such as lawyers and judges, use precedents (previously decided court cases) to support their arguments and help achieve their desired outcome in the present case \cite{Shulayeva2017}. Even in the Civil Law System (like in China and France), where legal arguments primarily rely on statutes, PCR remains essential as it offers key reference details, including the relevant statutes for past cases and the court's rulings, serving both legal experts and those seeking legal advice \cite{10.1145/3626772.3657887}.


While the general domain IR systems have progressed into neural retrieval models, PCR systems still rely largely on traditional and term-based methods such as \texttt{BM25} \cite{robertson2009probabilistic}. Researchers participating in COLIEE have demonstrated that \texttt{BM25} serves as a strong baseline, and most top-performing systems have employed models based on \texttt{BM25} combined with other techniques, such as TF-IDF and XG-Boost \cite{joshi-etal-2023-ucreat}. For example, a traditional language modelling approach \cite{10.1145/290941.291008} proposed in 1998 won the first place \cite{Ma2021RetrievingLC}, and a vanilla \texttt{BM25} got second place \cite{Yes-BM25} in COLIEE 2021 \cite{COLIEE2021}. Furthermore, many researchers such as \citet{DESIRES} and \citet{joshi-etal-2023-ucreat}, show that \texttt{BM25} outperforms many supervised transformer-based retrieval approaches. 

The limited success of neural retrieval models in PCR can be attributed to two main reasons.
\begin{enumerate}[label=\textbf{\textit{\arabic*.}}, wide, labelindent=0pt]
\vspace{-2mm}
\item \textit{\textbf{Long court cases}} - The court cases are lengthy in nature. Many state-of-the-art neural retrieval models, because of their reliance on \texttt{BERT} models \cite{devlin-etal-2019-bert}, have a context limit of 512 tokens \cite{10.1145/3397271.3401075,ren-etal-2021-rocketqav2}. While researchers have attempted to apply these neural retrieval models to PCR using techniques such as truncating court cases, they often result in information loss and suboptimal results \cite{DESIRES,joshi-etal-2023-ucreat,JNLP, premasiri-etal-2023-model}. 

\item \textit{\textbf{Lack of training data}} - The neural retrievals are supervised machine learning models and require a large number of training instances \cite{10.1145/3397271.3401075,ren-etal-2021-rocketqav2}. While several large PCR datasets exist, they are limited to a few languages and courts, and it is not always possible to find PCR datasets that are large enough to properly train neural retrievals, resulting in reduced performance. 
\end{enumerate}

The recently emerged LLM-based text embedders \cite{muennighoff2022sgpt,lee2024nv} can simultaneously address both of these challenges in PCR. First, they take up to 32,000 tokens as input, which is under the length of most court cases, addressing the first challenge we discussed. Secondly, as we will discuss in Section \ref{subsec:modelling}, LLM-based text embedders can be utilised in an unsupervised manner in the IR tasks, eliminating the need for model training and addressing the second challenge \cite{musts}. Furthermore, \citet{ni-etal-2022-large} demonstrate that LLM-based encoders also exhibit superior generalisability, showing significant improvements not only in the specific targeted scenario but also across a range of general tasks outside the fine-tuned domain. However, previous studies have not evaluated LLM-based text embedders in PCR benchmarks. In this research, we address this gap by answering the following two research questions (RQs). 

\noindent \textbf{RQ1} - How do the state-of-the-art LLM-based text embedders perform in different PCR benchmarks? 

\noindent \textbf{RQ2} - How well does the model ranking in the \texttt{MTEB} \cite{muennighoff-etal-2023-mteb} benchmark generalise to PCR benchmarks?

Answering these questions, in this paper, (1) we provide the \textbf{first comprehensive evaluation} of state-of-the-art LLM-based text embedders on PCR. (2) We show that our simple adaptation of LLM-based embedders \textbf{outperforms \texttt{BM25} and transformer-based methods} in PCR datasets across multiple languages and jurisdictions. We release the model code and evaluation scripts for the purpose of research usage via GitHub\footnote{Available at \url{https://github.com/DamithDR/case-retrieval.git}}

\section{Related work}

With the increasing volume of cases, there is a growing demand for automatic precedent retrieval systems to assist practitioners by providing prior cases relevant to the current case \cite{10.1007/s10506-015-9162-1}. Therefore, PCR has remained an active area of research in the IR community \cite{10.5555/2168120.2168126,10.1145/3626772.3657650,10.1007/978-3-031-56060-6_6,10.1145/3626772.3657693}. Several datasets have been released for the PCR task such as  \texttt{LeCaRDv2} \cite{10.1145/3626772.3657887}, \texttt{C3RD} \cite{10.1145/3589334.3645349} and \texttt{MUSER} \cite{li2023muser} for Chinese courts, \texttt{IL-PCR} \cite{joshi-etal-2023-ucreat} for Indian courts, \texttt{GerDaLIR} \cite{wrzalik-krechel-2021-gerdalir} and \texttt{LePaRD} \cite{mahari-etal-2024-lepard}  for United State's courts. Recent shared tasks, such as COLIEE \cite{goebel2024overview,goebel2023summary,kim2022coliee} and AILA \cite{parikh2021aila,bhattacharya2019fire} have facilitated the development of many PCR datasets. 

SAILER \cite{10.1145/3539618.3591761} introduced a structure-aware pre-trained model for the prior case retrieval task. Utilising encoder-decoder architecture, SAILER proposes a fact encoder, a reasoning decoder and a decision decoder, which are pre-trained on Chinese and United States case law. 
CaseLink \cite{10.1145/3626772.3657693} introduces a different approach for PCR by creating a Global Case Graph. They utilise semantic and legal charge relationships in addition to the reference relationships to populate the case graph. However, evaluating similarity among case law is a complex task. DELTA \cite{li2025delta} proposed an encoder-based pre-trained model with structural word alignment to methodically align the relevant facts closer and the irrelevant ones distant. Moreover, PCR has also been studied at the paragraph level. \citet{t-y-s-s-etal-2024-query} created a specific dataset for paragraph retrieval in the European Court of Human Rights (ECtHR) and performed zero-shot and fine-tuned experiments with multiple encoder models.  BERT-PLI \cite{ijcai2020p484} fine-tunes a BERT model for the sentence pair classification task and utilises the semantic relationships to calculate the relevance prediction using an interaction map. 

Given the importance of information retrieval in the legal domain, multiple research events have been organised in this area. Few notable events are LIRAI \cite{10.1145/3603163.3610575} workshop focused on information retrieval systems generally in the legal domain, and particularly in case law COLIEE \cite{10.1007/978-981-97-3076-6-8} shared-task focuses on case law retrieval systems. The results of the competition suggest that BM25-like retrieval models are still effective in the context of lengthy text, as in case law.

\section{Methodology}

\subsection{Data}
Considering the diversity of jurisdictions and languages, we selected four popular PCR datasets: \texttt{IL-PCR} \cite{joshi-etal-2023-ucreat}, \texttt{COLIEE-2022} \cite{10.1007/978-3-031-29168-5_4},  \texttt{MUSER} \cite{10.1145/3583780.3615125}, and \texttt{IRLeD} \cite{mandal2017overview}. We further summarise the details of the datasets in Table \ref{tab:datasets_table}. We used the test set (both candidates and queries) of \texttt{IL-PCR}, test queries in the \texttt{COLIEE-2022} and the whole dataset of \texttt{MUSER} and \texttt{IRLeD}, as there were no separate splits.

\begin{algorithm}
\small
\caption{Ranking Court Cases using Precomputed LLM Embeddings and MAP Evaluation for Multiple Datasets}
\begin{algorithmic}[1]
\Require Dataset collections $\mathcal{D} = \{D_1, D_2, \dots, D_5\}$ (court cases), Embedding model $M$
\Require Queries $Q = \{q_1, q_2, \dots, q_m\}$ for each dataset in $\mathcal{D}$

\State Precompute embeddings $e_q$ for each query $q \in Q$ using $M$
\For{each dataset $D \in \mathcal{D}$}
    \State Precompute embeddings $e_c$ for each candidate case $c \in D$ using $M$
    
    \For{each query $q \in Q$}
        \For{each candidate case $c \in D$}
            \State Calculate cosine similarity $s(q, c) = \frac{e_q \cdot e_c}{\|e_q\| \|e_c\|}$
            \State Append $(c, s(q, c))$ to list of candidates for $q$
        \EndFor
        
        \State Sort candidates by descending similarity scores
        \State $R_q \gets \text{ranked list of candidates for } q$ in dataset $D$

        \For{$k = 1$ to Top-$K$ candidates in $R_q$}
            \State Calculate Precision@$k$, Recall@$k$, and F-score@$k$
        \EndFor
        
    \EndFor
    \State Calculate Mean Average Precision (MAP) across all queries in dataset $D$
\EndFor

\end{algorithmic}
\label{algo:retrieval}
\end{algorithm}

\begin{table*}[ht]
    \centering
    \small
    \scalebox{0.96}{
    \begin{tabular}{|c|c|c|c|c|}
        \cline{2-5}
        \multicolumn{1}{c|}{} & \makecell[c]{\texttt{IL-PCR} \\  \cite{joshi-etal-2023-ucreat}} & \makecell[c]{\texttt{COLIEE} \\ \texttt{2022} \cite{10.1007/978-3-031-29168-5_4}} & \makecell[c]{ \texttt{MUSER} \\ \cite{10.1145/3583780.3615125}} & \makecell[c]{ \texttt{IRLeD} \\ \cite{mandal2017overview}} \\
        \hline
        Total No of queries & 237 & 300 & 100 & 200 \\
        \hline
        Total no of candidates & 1727 & 1263 & 1038 & 2000 \\
        \hline
        Average no of words in queries & 6766.32 & 5107.03 & 1993.12 & 7801.15 \\
        \hline
        Average no of words in candidates & 7046.36 & 4700.66 & 1747.52 & 7294.31 \\
        \hline
        Language & English & English & Chinese & English \\
        \hline
        Jurisdiction & India & Canada & China & India \\
        \hline
    \end{tabular}
    }
    \caption{Details of the datasets}
    \label{tab:datasets_table}
\end{table*}

\subsection{Modelling}
\label{subsec:modelling}
As we mentioned earlier, we utilise LLM-based embedders in our research to investigate prior case retrieval across different jurisdictions. We choose three embedding models which are among the top five in \texttt{MTEB} \cite{muennighoff-etal-2023-mteb} leaderboard\footnote{Available at \url{https://huggingface.co/spaces/mteb/leaderboard}} as of October 2024\footnote{\texttt{NV-EMBED}\cite{lee2024nv} is the best model in the MTEB leaderboard as of October 2024. However, we were unable to use the model on an NVidia L40 48G GPU. Therefore, we only used the models ranked 2nd, 3rd and 4th in MTEB.}. Namely they are; \texttt{BAAI\//bge-en-icl}\footnote{Available at \url{https://huggingface.co/BAAI/bge-en-icl}} \cite{li2024makingtextembeddersfewshot}, \texttt{Salesforce\//SFR-Embedding-2\_R}\footnote{Available at \url{https://huggingface.co/Salesforce/SFR-Embedding-2\_R}} \cite{SFR-embedding-2} and \texttt{dunzhang\//stella\_en\_1.5B\_v5}\footnote{Available at \url{https://huggingface.co/dunzhang/stella\_en\_1.5B\_v5}}. We used Salesforce\//SFR-Embedding-2\_R and dunzhang\//stella\_en\_1.5B\_v5 in combination with SentenceTransformer Python package \cite{reimers-gurevych-2019-sentence} while \texttt{BAAI\//bge-en-icl} with FlagEmbedding Python package.

Following our motivation to address the challenge of the lengthiness of prior cases, LLM-based embedders help us to obtain a high-dimensional vector representation of each case. Unlike transformer-based models, these models support a high context length. As the models can have a high context length, they can capture most of the information in lengthy case documents. The first step is to obtain embeddings for each candidate and query case for all datasets separately, utilising the above-mentioned models.

For the retrieval process, we iterate over each query case embedding and calculate the cosine similarity with all candidate embeddings. We rank the most similar embeddings with a higher rank and the least similar ones with a lower rank. We use cosine similarity as our primary metric for calculating the similarity. Our retrieval algorithm is shown in Algorithm \ref{algo:retrieval}.

We calculate the Mean Average Precision (MAP) as our primary evaluation metric. We also calculate precision@k, recall@k and F score@k values where k=\{1, 5, 10, 15...50 and 100\} following the recent research in PCR \cite{joshi-etal-2023-ucreat}. We use two baselines. First, we employ BM25 \cite{robertson2009probabilistic}, which is a strong and popular baseline for PCR, as we mentioned before. As the second baseline, we employ sentence-transformer with \texttt{LEGAL-BERT}. We trained a \texttt{LEGAL-BERT} model using the training set of the \texttt{IL-PCR} dataset. We used the positive samples from the dataset and added five negative samples for each query case to create the training data. The model was trained with learning\_rate=2e-5, epoch=1, batch\_size=16. The resulting model was used to create embeddings following the same method as other models to evaluate them. 

\begin{table*}[ht]
    \centering
    \small
    \scalebox{1.08}{
    \begin{tabular}{|c|c|c|c|c|c|c|c|c|c|c|c|c|}
    \hline
    \multirow{2}{*}{Model} & \multicolumn{3}{c|}{\texttt{IL-PCR}} & \multicolumn{3}{c|}{\texttt{COLIEE}} & \multicolumn{3}{c|}{\texttt{IRLeD}} & \multicolumn{3}{c|}{\texttt{MUSER}} \\
    \cline{2-13}
         & MAP & F & k & MAP & F & k & MAP & F & k & MAP & F & k \\
         \hline
         \texttt{bge-en-icl} & 0.42 & 0.31 & 5 & 0.29 & 0.22 & 5 & 0.25 & 0.25 & 5 & 0.10 & 0.08 & 25 \\
         \texttt{SFR-Embedding-2\_R} & \textbf{0.47} & \textbf{0.34} & 5 & \textbf{0.32} & \textbf{0.25} & 5 & \textbf{0.27} & \textbf{0.27} & 5 & 0.12 & 0.10 & 25 \\
         \texttt{stella\_en\_1.5B\_v5} & 0.44 & 0.32 & 5 & 0.32 & 0.24 & 5 & 0.26 & 0.25 & 5 & \textbf{0.14} & \textbf{0.11} & 25 \\
         \hline
         \texttt{LEGAL-BERT} & \cellcolor{blue!25} 0.27 & \cellcolor{blue!25} 0.19 & \cellcolor{blue!25} 5 & 0.14 & 0.11 & 5 & 0.09 & 0.08 & 10 & 0.04 & 0.04 & 10 \\
         \texttt{BM25} & 0.16 & 0.18 & 10 & 0.26 & 0.20 & 5 & 0.20 & 0.20 & 5 & 0.12 & 0.10 & 25 \\
         \hline
    \end{tabular}
    }
    \caption{Model performance on different datasets. Column Model shows the
model used in the experiment and columns \texttt{IL-PCR}, \texttt{COLIEE}, \texttt{IRLeD}, \texttt{MUSER} show the dataset used for the
experiment. Column MAP shows the mean average precision results, and column F score shows the best F score achieved by the model for the dataset. Column k value indicates the corresponding k value to the best F score. The only instance where we used a supervised model is coloured in blue.}
    \label{tab:results_table}
\end{table*}

\begin{figure*}[h!]
    \centering
    \begin{subfigure}{0.45\textwidth}
        \centering
        \includegraphics[width=\textwidth]{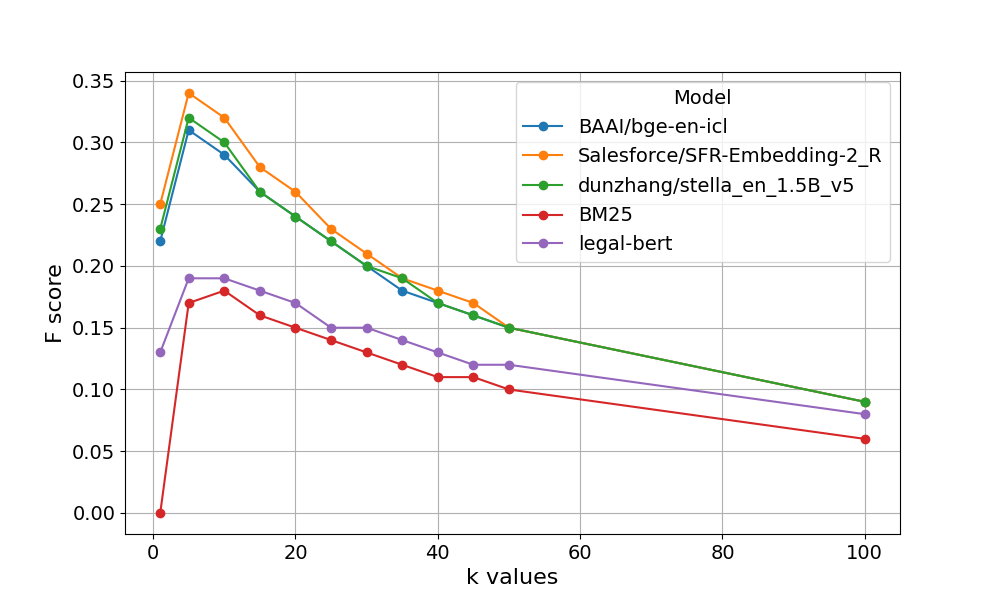}
        \caption{\texttt{IL-PCR}}
        \label{fig:ilpcr_f}
    \end{subfigure}
    \hfill
    \begin{subfigure}{0.45\textwidth}
        \centering
        \includegraphics[width=\textwidth]{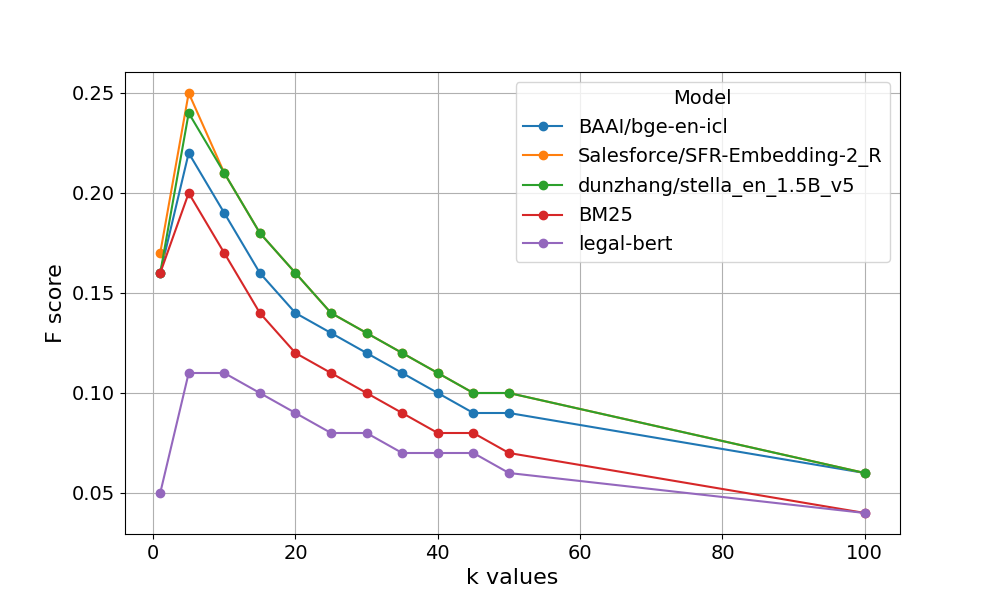}
        \caption{COLIEE}
        \label{fig:coliee_f}
    \end{subfigure}
    
    \vspace{1em} 

    \begin{subfigure}{0.45\textwidth}
        \centering
        \includegraphics[width=\textwidth]{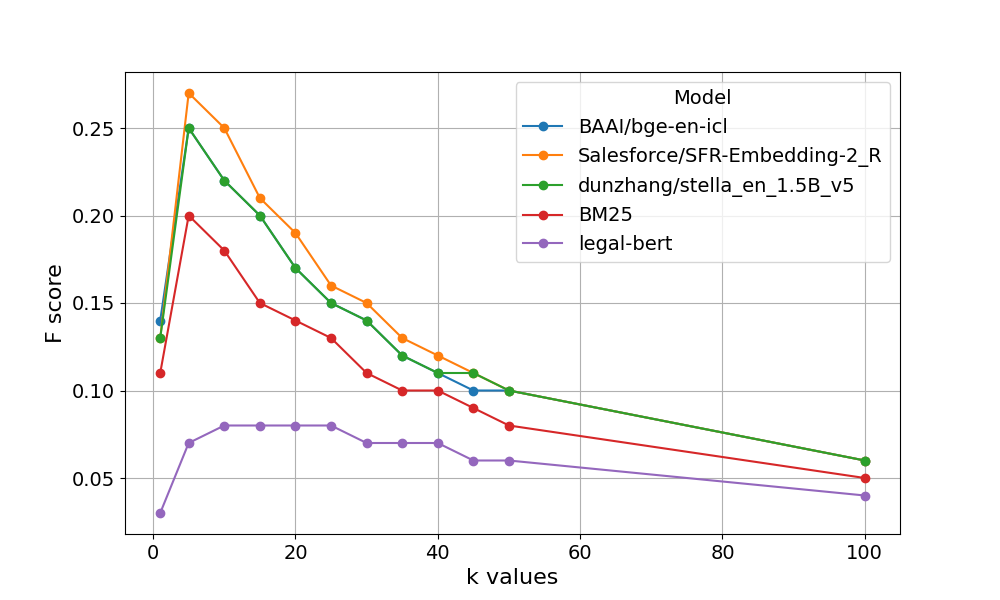}
        \caption{IRLeD}
        \label{fig:irled_f}
    \end{subfigure}
    \hfill
    \begin{subfigure}{0.45\textwidth}
        \centering
        \includegraphics[width=\textwidth]{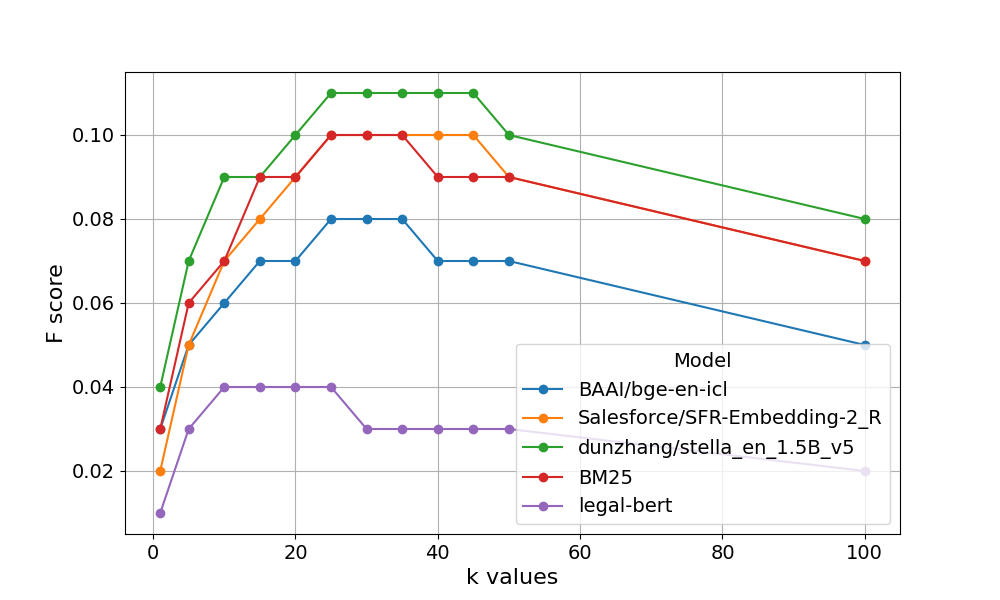}
        \caption{MUSER}
        \label{fig:muser_f}
    \end{subfigure}

    \caption{Change of F score with respect to k values.}
    \label{fig:main_figure}
\end{figure*}

\section{Results and Analysis}
Table \ref{tab:results_table} summarises the MAP values and best F scores achieved by each model and respective k values. Figure \ref{fig:main_figure} illustrates the F-score curves for all datasets using all models. 

As can be seen in the results, LLM-based embedders outperform the \texttt{BM25} baseline with a clear margin in all datasets, answering our \textbf{RQ1}. As can be seen, for all the k values, LLMs perform better than \texttt{BM25}, showing their effectiveness. \texttt{IL-PCR} dataset shows a 0.16 improvement in F score compared to \texttt{BM25} in \texttt{SFR-Embedding-2\_R model} while reporting 0.47 MAP score. In \texttt{COLIEE} dataset, both \texttt{SFR-Embedding-2\_R model} \texttt{stella\_en\_1.5B\_v5} and show 0.06 improvement in MAP compared to \texttt{BM25}. Furthermore, in \texttt{IL-PCR}, all LLM-based encoders outperform the supervised legal-bert model. In other datasets, too, LLM-based embedders outperform the legal-bert model, showing that they generalise well compared to other unsupervised models. Overall, it is clear that LLM-based embedders provide a promising solution to PCR.

From the LLMs \texttt{SFR-Embedding-2\_R} shows the best performance for three out of four datasets from both MAP and F scores, while \texttt{stella\_en\_1.5B\_v5} is the best performer for the \texttt{MUSER} dataset. However, it should be noted that \texttt{bge-en-icl} is the best model out of these models in the MTEB benchmark, yet it does not outperform other models in the PCR tasks. With this finding, we answer \textbf{RQ2}, the model ranking in the \texttt{MTEB} benchmark does not generalise into the PCR benchmarks. While \texttt{MTEB} benchmark contains IR tasks, it does not contain any PCR tasks, which explains our observation to \textbf{RQ2}.

\section{Conclusion}

In this paper, we empirically showed that state-of-the-art LLM-based embedding models in \texttt{MTEB} benchmark outperform \texttt{BM25} in multiple PCR datasets in multiple jurisdictions. However, as \texttt{MTEB} does not contain any PCR tasks, the model ranking in \texttt{MTEB} is not reflected in PCR datasets. Overall, LLM-based embedding models provided better results in all the PCR datasets, outperforming popular baselines, \texttt{BM25} and other supervised baselines.

As the first comprehensive evaluation of LLM-based embedding models in the PCR task, this research will open several future research directions. First, the IR community needs to incorporate PCR datasets widely into IR benchmarks. Secondly, LLM-based embedders should be trained in PCR tasks so that they will provide better results than the unsupervised approach.

\section*{Acknowledgements}
We would like to thank the anonymous reviewers for their positive and valuable feedback. We further thank the creators of the datasets used in this paper for
making the datasets publicly available for our research.

The experiments in this paper were conducted in UCREL-HEX \cite{UcrelHex}. We would like to thank John Vidler for the continuous support and maintenance of the UCREL-HEX infrastructure, which enabled the efficient execution of our experiments.

\section*{Limitations}
We acknowledge that there are bigger models with higher-dimensional representations; however, we do not conduct experiments on these models due to hardware resource limitations.

We used cosine similarity as our only relevancy metric; however, there are other methods that can be explored to get the similarity between two vectors. Furthermore, our experiments are currently limited to a few jurisdictions and languages, such as English and Chinese, which we plan to expand in the future, as well as developing specific models for PCR by further pre-training on PCR datasets.



\bibliographystyle{acl_natbib}
\bibliography{anthology,ranlp2025}


\end{document}